\documentclass[11pt]{article}

\usepackage[a4paper,margin=4cm]{geometry}

\usepackage[utf8]{inputenc}
\usepackage[T1]{fontenc}
\usepackage{xspace} 
\usepackage{framed}

\usepackage{textcomp}
\usepackage{hyperref}
\usepackage{tabularx}
\usepackage{booktabs}
\usepackage{multirow}
\usepackage{charter}

\usepackage{listings}
\usepackage{fancyhdr}
\usepackage{graphicx}
\usepackage{eurosym}
\usepackage{titlesec} 
\usepackage{marginnote}
\usepackage{enumitem}
\usepackage{comment}

\usepackage{pdfpages}
\usepackage[doi=false,url=false]{biblatex}
\bibliography{biblio,monperrus}

\title{Software that Learns from its Own Failures}
\author{Martin Monperrus\\ University of Lille \& Inria\\martin.monperrus@univ-lille1.fr}

\begin{document}

\maketitle

\begin{abstract}

All non-trivial software systems suffer from unanticipated production failures. However, those systems are passive with respect to failures and do not take advantage of them in order to improve their future behavior: they simply wait for them to happen and  trigger hard-coded failure recovery strategies. Instead, I propose a new paradigm in which software systems learn from their own failures. By using an advanced monitoring system they have a constant awareness of their own state and health. They are designed in order to automatically explore alternative recovery strategies inferred from past successful and failed executions. Their recovery capabilities are assessed by self-injection of controlled failures; this process produces knowledge in prevision of future unanticipated failures.

\end{abstract}

\section{Introduction}

\newcommand{\wsection}[1]{\vspace{.2cm}\textbf{#1}}

All non-trivial software systems suffer from unanticipated production failures.
For instance, on March 12 2012, the
Mozilla Firefox web browser has crashed 270455 times.
A software failure is commonly defined as an output that does not correspond to the user's expectations \cite{avizienis2004basic}. In the Mozilla Firefox example,  the failure is that the program closes (actual behavior) instead of rendering the requested web page (expected behavior). A failure is caused by a fault in the code  or an unexpected environmental condition \cite{avizienis2004basic}. 
The classical view on software failures is to combat them with techniques for:
detecting faults using static and dynamic software analysis; 
proving the absence of certain faults; 
and
improving the development processes and tools to prevent the introduction of faults. 
Those research threads yielded fundamental advances in software engineering but failed to eradicate software failures.
In this paper, I take a ground-breaking perspective on this topic: instead of being passive with respect to production failures, software systems should actively monitor, exploit and learn from them.

\wsection{The Computing World Today}
Let us take the story of Pat, working on her manuscript using a text processing application. On June 26, 2014, a software failure in her text processing document results in complete loss of the last four weeks of writing.
Pat's failure is useless: the conditions in which the failure happened are unknown, the fact that Pat's failure also happened on Bob and Alice's machines is not shared, the reason for which the recovery planned by the developer failed is lost.

\wsection{The Computing World Tomorrow}
All text processors of the same version would collect and share execution information. When Pat's failure occurs, detailed information is sent back to an observation server running diagnosis algorithms. The server identifies a key similarity between 17 failures. The diagnosis algorithms analyze the data and triggers fine-grain monitoring. After having collected 5 more failures and the associated detailed execution information, the server runs a learning algorithm that identifies that the code variable ``plugin'' is responsible for the failure.
The server then synthesizes a code change as a solution. To validate the change, the system proactively injects the same failure on Alice's machine and validates that the inferred solution avoids the severe data loss. The failure of user Pat is now part of in a global adaptive and collaborative learning process, the software system (the text processor) gets better after the occurrence of the failure. Pat's failure has become useful.

\textbf{My vision is that software systems can learn from their own failures, as humans do}.
By ``learning'', self-identifying execution and environmental patterns in which the failures occur and self-synthesizing memory and code changes so that the failures become harmless. 
Learning is an active process: software systems must constantly assess whether the learned healing capabilities succeed, must proactively explore alternative recovery strategies, for instance using self-injection of failures.

Recent research suggests that automatically fixing software failures is possible \cite{Monperrus2014}. Weimer et al. \cite{Weimer2010}, Zeller et al. \cite{Dallmeier2009} and others have proposed algorithms that generate valid patches to automatically repair failure scenarios. Those systems have a fundamental limitation: they work offline and not in production.

To realize my vision, I propose three research directions:

\begin{itemize}
\item Designing new adaptive and collaborative software monitoring systems. They will provide the input data to the learning and inference algorithms, about the software executions and failures across multiple machines.

\item  Inventing resourceful dynamic repair techniques that learn to self-improve. They will employ specification mining and code synthesis for achieving runtime adaptive failure recovery and exploring the recovery space.

\item Characterizing proactive perturbation of software systems with injected failures. When failures become useful, one can trigger controlled failures in production for the system to learn and improve its runtime knowledge and consciousness of its environment and recovery capabilities.

\end{itemize}

\section{State of the Art}

The vision is close to the research on self-healing software \cite{koopman2003elements,keromytis2007characterizing}, which has also been referred to as software immune systems \cite{sidiroglou2005building}. 
Now, the literature seems to prefer the term ``automatic repair'' (runtime repair, dynamic repair) \cite{Monperrus2014}.

Research on automatic software repair and self-healing systems has started in years 2000.
It can be decomposed in two currents \cite{Monperrus2014}.
The first is called ``behavioral repair'' concerns repair of the system code.
A famous technique is Genprog \cite{Weimer2010}. It consists of generating a patch so as to  make a failing test case passing. It employs different code manipulation techniques and a kind of genetic optimization to drive the repair process. Very recent advances in this active field include DirectFix \cite{directfix}.
Genprog and Directfix run offline, they do not consider production failures and only work on failures for which one failing test case has been written. 
While the code manipulation techniques of Genprog are powerful, it misses the essential capability to be executed in production.

The second current of automatic software repair consists of changing the execution state at runtime. 
As early as 1980, Taylor and colleagues \cite{taylor1980redundancy} introduced ``robust data structures'' which are able to repair their own state at runtime. More recently, Demsky and Rinard  \cite{demsky2003automatic} proposed a similar approach for data structure repair \cite{demsky2003automatic}.
Also much related, Perkins et al.'s ClearView  \cite{Perkins2009} is a production system. Plugged into legacy x86 software, it mines runtime invariants on CPU registers and restores those invariants upon unanticipated failures. This is a strong limitation: previous experiments with reasoning on unanticipated production failures \cite{cornu:hal-01062969} 
has shown that real-life recovery goes far beyond invariants on CPU registers.

The closest research on this topic is by Sidoroglou et al. \cite{sidiroglou2009assure}. They have devised a system, called Assure, which is able to recover from unanticipated failures using a technique called ``error virtualization''. Error virtualization consists of repurposing failure handling-code for a larger class of failures. For instance, if some code has been written for handling the case where a file does not exist, it can be repurposed for handling the case where the file is not readable. 
Assure aims at handling unanticipated failures but it considers all failures in isolation.
What is missing in Assure is to link failures together, to reason about the fact that many failures have the same cause, to understand why some recovery attempts succeed while others fail. 

More fundamentally, all those systems are passive with respect to failures, they just wait for failures to happen. I propose to explore a novel direction: software systems must become active, as in ``active learning''. They can be active with respect to failures in a number of promising manners: by self-injecting failures to self-assess the recovery capability and by exploring alternative recovery strategies to explore the recovery space. 

Fault injection can be casted in a broader concept of execution perturbation. 
Ammann and Knight's ``data diversity'' \cite{ammann88} consists of changing values at runtime so as to complete a computation in the presence of failures. 
Tang and colleagues showed that execution perturbation of floating point programs is an efficient technique to uncover failures \cite{tang2010perturbing}. 
However, nobody has ever studied automatic execution perturbation in production for constantly assessing recovery capabilities and exploring the recovery space.

\section{Research Agenda}

To construct software that learns from its own failures, 
there are three axes to take into account.
First, software must become deeply aware of its own state and health. This can be achieved through novel automated monitoring systems.
Second, software must be capable of changing and synthesizing its own recovery code at runtime. This requires completely revisiting the way recovery is  handled.
Third, once software has the capability to improve upon failures, it becomes ready to proactively explore its reactions to new failures, to proactively explore alternative new recovery strategies.

\subsection{Future of Monitoring}

\begin{figure}
\caption{At runtime in production, the software system is augmented with a collaborative monitoring module, a resourceful recovery module and a perturbation module. All together, they form \textbf{software that learns from its own failures}.}
\label{fig-overview-runtime.pdf}
\centering\includegraphics[scale=.6]{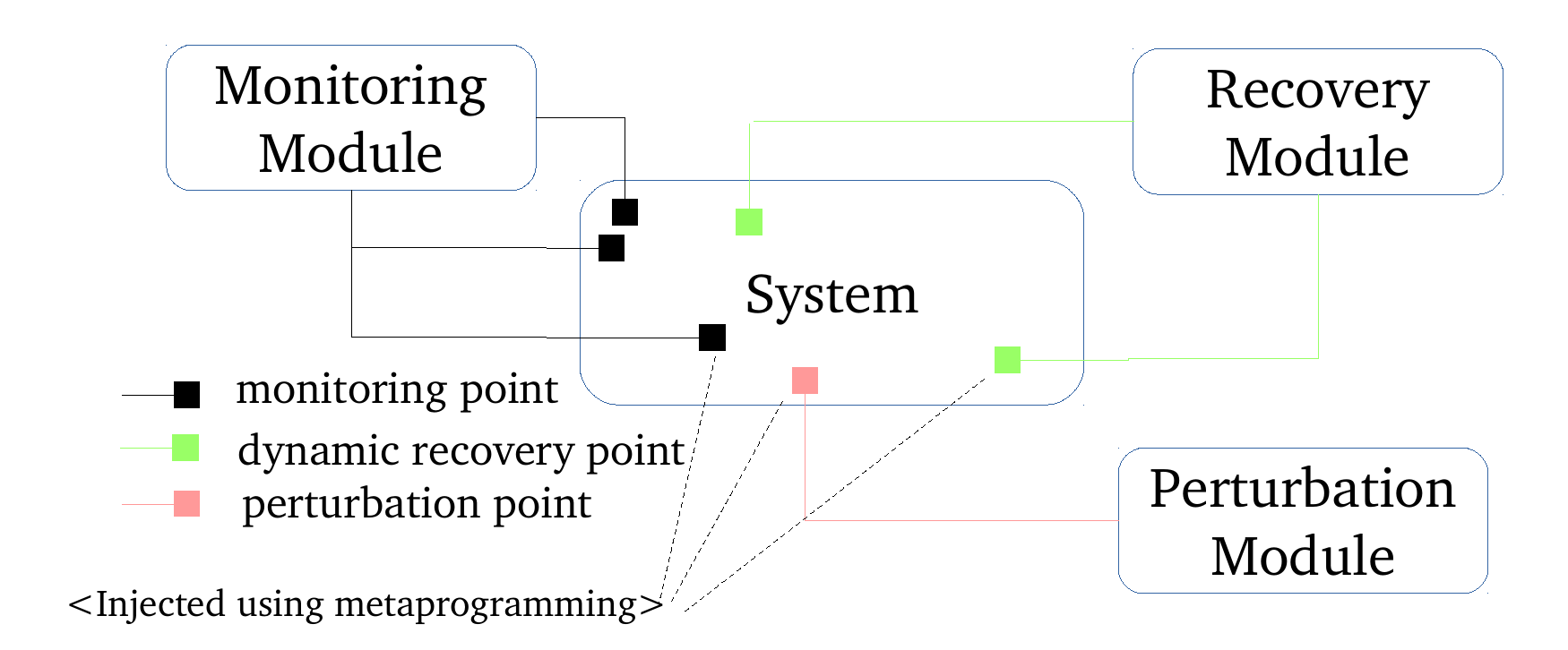}
\end{figure}

Software that learns from its own failures needs to reason on its own internal state, health and history.
To do so, it needs to embed a monitoring system.
However, today's monitoring systems are not able to give an accurate view of  the execution state. Let's consider the following piece of code:

\begin{lstlisting}
error = login(user, password, encryptionMethod);
if (error==true) {
  log("error during login for user:"+ user);
}
\end{lstlisting}

The call to function \texttt{log} is a monitoring probe, it has been manually added by a conscientious developer to keep a trace of login errors. Now assume that legitimate user ``Bob'' is not able to login due to a rare failure. The failure is due to the encryption method that is used. 
To be able to diagnose the failure, the system must understand that when the encryption method is ``md5'' it succeeds and when it is ``sha1'', it fails. 
To do so the monitoring probe must look like the following, which has not been foreseen by the developer \texttt{log("error during login: "+context(user, encryptionMethod));
}.

Upon failures, the monitoring system must collect all relevant information. One cannot only rely on manually added monitoring directives, because they are doomed to be incomplete:
either calls to the monitoring system are missing, or not all variables are collected, as illustrated in the above example.

One can employ automatic injection of monitoring probes through meta\-programming, as shown in Figure \ref{fig-overview-runtime.pdf}.
A monitoring system for software that learn from its failures has to meet the following requirements.
First, it collects very fine grain monitoring data, up to the level of variables. This is possible because metaprogramming gives access to all elements of the code.
Second, the monitoring system shall be adaptive. For instance, in nominal mode, it only collects the value of variable ``user'' in order to keep trace of all connections in the system. Upon the first failure, it starts to collect the other variables that are involved in the login process, so as to diagnose that the failure is due to the wrong value of ``encryptionMethod''.
Third, the cost of monitoring (memory, disk, bandwidth) is dynamically controlled. For instance, if disk space is limited, the monitoring system decides to only collect one out of two calls to function ``login''. This is a tradeoff between keeping the core functionalities up and the speed of diagnosis.
Fourth, monitoring must be collaborative. The same failure happens on many machines running the same software application. The failure information on machine  $x$ can be cross-compared with the same failure happening on machine $y$. The comparison of similarities and  differences drives the diagnostic.

\begin{framed}
I envision a monitoring system that is adaptive, collaborative and resource-aware. Nobody has ever put those three capabilities together.
\end{framed}

When monitoring becomes fine-grained and collaborative, it may become a threat to the privacy of the application's user. Special care must be taken so as to protect the end users' privacy. Anonymization and sampling have to be used so that no malicious attacker can exploit the system to track, spy or steal the users.

\noindent\textbf{Barriers: } The resources taken by monitoring compete with the ones needed for the code functionalities  (disk, memory and bandwidth). 
Also, there is a major size issue: there are thousands of machines that cooperate to share monitoring data. 
This requires an architectural blueprint as well as core optimizations that are beyond the state of the art.

\subsection{Acting upon Unanticipated Failures}

In today's software systems, the mainstream way to communicate and handle failures is exception handling. In software design books, entire chapters are dedicated to discussing good and bad practices on exception handling \cite{cwalina2008framework}. 
For instance, the Hadoop distributed filesystem does most of its failure-handling using exceptions through 9888 catch block.
Briefly, classical exception handling works as follows:

\begin{minipage}{7cm}
\begin{lstlisting}
// failure detection
if (file==null) { // failure condition
  throw new InputException();
}
\end{lstlisting}
\end{minipage}

\begin{minipage}{6.5cm}
\begin{lstlisting}
// failure handling
catch (InputException) { // recovery condition
  defaultText = "foo bar"; // recovery strategy
}
\end{lstlisting}
\end{minipage}\\

One sees the three main components of failure handling using exceptions:
1) the failure condition (here \texttt{file==null}) encodes a predicate on the system state whose value encodes the presence of a failure;
2) the recovery condition states when a failure can be handled (here when an exception of type \texttt{InputException} arrives at this location);
3) the recovery code repairs the system state it is the content of the catch block (here \texttt{defaultText = "foo bar"}).
Today, all those three components are manually written and hard-coded. This is a fundamental limitation.

In today's open-ended systems, the kinds of failures and recovery are open-ended.
As said above, the Eclipse development environment runs on at least 5 millions different machines. Can the developer foresee all possible failures? Can she hard-code all possible failure conditions, recovery conditions and recovery strategies? No.

I claim that that all those three components must become adaptable instead of being manually hard coded, as shown in the following example:

\begin{minipage}{6.5cm}
\begin{lstlisting}
// failure detection
if (failureDetected(file, user)) {
  throw new InputException();
}
\end{lstlisting}
\end{minipage}

\begin{minipage}{7cm}
\begin{lstlisting}
// failure handling
catch (failureRecoverable(exception, context)) { 
  recover(defaultText,output);
}
\end{lstlisting}
\end{minipage}\\
In this simplifying example, the main changes compared to the previous one is that the failure detection condition becomes the result of the evaluation of a function, so do the recovery condition and the recovery code. 
By replacing an hard-coded condition with a function evaluation, the failure-handling can be become resourceful and capable of handling unanticipated failures. In this context, the notion of ``resourceful'' is close to that of Abbott \cite{abbott1990resourceful} and means two things:
first, resourceful failure-handling code can be changed at runtime if previous evaluations failed to heal the system, second it can act upon an arbitrary number variables of the system, incl. those that have not been foreseen by the developers. 
In the example, although the manually written failure condition only involves variable ``file'', the correct failure condition detection may actually involve variable ``user''.
This point also holds for the recovery condition and the recovery code since the developer cannot perfectly foresee the variables required for recovery in all cases.

This vision poses a number of challenges. First, it must be non-conflicting with the existing failure handling code (the recovery must gracefully augment the 9888 existing catch blocks in the case of Hadoop). 
Second, it must work in intimate collaboration with a monitoring system so as to reason on when and how a resourceful algorithm must be run to synthesize a new failure condition, recovery condition or recovery strategy. Third, it must be compatible with existing compile-time and runtime support for exceptions (not all languages support parameterizing the catch exception by function evaluation as described in our example).

One enabling solution is the introduction of runtime evaluation and first class execution objects at all stages of the failure-handling process:
first failure recovery strategies can then be reused even in unanticipated cases (as opposed to only triggered in the specified cases);
second failure detection recovery can be adapted: when recovery fails, alternative solutions in a recovery space are explored, for instance by taking into account new portions of the system state (new variables).
Last but not least, the recovery itself can be seen declaratively:
the call to \texttt{recover(defaultText,output)} searches for a solution to a recovery problem, with a declarative specification of the recovery problem (given a failure detection condition, a recovery condition and a necessary recovery post-condition). 

\begin{framed}
I propose to consider recovery under three novel angles: recovery should be the result of a synthesis problem; recovery should be a first-class execution runtime object with full intercession capabilities; recovery should become a search problem with alternative and competing solutions.
\end{framed}

\noindent\textbf{Barriers: } The main barrier for achieving resourceful recovry is semantics. As shown in our empirical study \cite{monperrus:hal-01093908}, recovery uses the full range of  programming language constructs with rich and complex semantics.
Also, the diversity of failure conditions and recovery strategies may be too large for keeping a complete record of all encountered failures and their solutions, so as to ensure the automatic exploration of new portions of the search space.

\subsection{Constant Assessment of Recovery Capabilities with Injected Perturbations}

Once one has changed the perspective on software failures, once one considers that software failures are an opportunity for the system to self-improve, this opens radically new research directions.
Today's research devises systems that are passive with respect to failures in production: they only handle failures that ``naturally'' happen, where ``naturally'' means being triggered by an external and uncontrolled cause. 
I envision systems that proactively perturb the execution with injected failures.
The injected failures are carefully qualified with respect to the expected recovery contracts.
Let us consider a concrete example.

According to our statistics on the Internet, the most common failures in Java software are null dereferences (``null pointer exceptions''). Null dereferences cause desktop, server and mobile applications to crash on a daily basis.
Now, let's consider a software system built using perturbationapplied to null dereferences.
The system embeds an advanced monitoring system as described above, as well as resourceful recovery for null dereferences.
As such, the system has already overcome 15 unhandled null dereferences.
Now, the system is augmented with a module that selectively injects null values in memory.
This system would inject $x$ (say 3) null values per day so as to 1) assess that the system does not crash upon null dereferences, and 2) validate   synthesized recovery pro-actively.

On the hardware side of production systems, this idea is already applied. For instance, datacenters are regularly subject to power cut so as to assess whether the alternative sources of power are up and running. I propose to explore this disruptive idea on the software side, to constantly assess whether the embedded software recovery code well handles software failures.

This can be seen as an application of the scientific method to failure recovery.
The scientific method states that all hypotheses must be experimentally validated using falsification experiments. 
A recovery capability is also an hypothesis: if an event of type $x$ happens, the system is able to survive. 
By injecting an event $x$ and assessing successful failure-recovery, one ensures the truthfulness of the recovery hypothesis in production.
In biology, ``hormesis'' refers to the positive response of biological systems (e.g. a cell) in response to a stressor. 
This can be seen as the exploration of the notion of hormesis in the domain of software systems. Hormesis is close to antifragility \cite{antifragile} and to this extent, this paper further explores the engineering of antifragile software \cite{Monperrus14}.

This is different and complementary from performing failure analysis in testing phase.
Failure analysis during testing enables to validate specific, well formed, small failures.
However, production systems are too big and too interconnected to be reproduced in a controlled testing environment. 
This results in unanticipatable situations and behaviors. Failure injection in production aims at tackling those unanticipatable, untestable failures that necessarily happen in production.

\begin{framed}
Software systems must become active with respect to software failures:
they must constantly assess their own recovery capabilities with failure injection,
they must  constantly explore the recovery space based on the injected failures.
The injected failures are carefully controlled to both maximize the knowledge gained from each injection and to mitigate their impact and cost.
\end{framed}

\noindent\textbf{Barriers: } 
This research direction has never been explored before. 
The idea of injecting faults in production is fundamentally disturbing. We do not know whether one can fully control the impact of injected faults, and whether the benefits obtained with failure injection (better failure recovery; improved knowledge) outperforms the losses (unfilled requests, unsatisfied users).

\section{Conclusion}

In this paper, I have sketched vision of software that learns to self-improve from its own failures.
To realize the vision, I set up a research agenda in three points: devising the next-generation of collaborative and adaptive monitoring systems, inventing a new generation of recovery based on code synthesis and automated exploration of the recovery space, and characterizing failure injection in production.
The last point is radically new and very promising: fault-injection in production is the only way to proactively assess the recovery capabilities and to proactively explore the recovery space.

\printbibliography

\end{document}